# Implementing an intrinsically integrated game on Newtonian mechanics in the classroom: outcomes in terms of conceptual understanding and transfer

Anne van der Linden[a], Ralph F. G. Meulenbroeks[b] and Wouter R. van Joolingen[c]

Freudenthal Institute, Utrecht University, Utrecht, The Netherlands

[a] a.vanderlinden@uu.nl, ORCID: 0000-0001-7970-0664
[b] r.f.g.meulenbroeks@uu.nl, ORCID: 0000-0001-6614-9156
[c] w.r.vanjoolingen@uu.nl, ORCID: 0000-0002-4271-2861

Corresponding author: Anne van der Linden. a.vanderlinden@uu.nl

**Abstract**

Digital educational games have demonstrated large variations in learning outcomes and transfer. Furthermore, educational games are usually embedded in a larger educational setting. This case study evaluates in detail a lesson around an educational game designed to foster transfer. The game, Newton's Race, is an intrinsically integrated game on Newtonian mechanics. The learning goal of the lesson is understanding and applying the relationship between forces and motion. Outside of the game, lesson activities include a debriefing session, a generalisation assignment, and an assignment on transfer situations. This lesson was evaluated using a mixed-methods approach. A pre- posttest design (N=27) demonstrated a large significant learning effect ($p = .002$, $d = .908$). Transfer, as measured within the posttest, was also fostered significantly. In the qualitative part of the study, students' written statements on the worksheets and students' utterances during the discussion were analyzed using open coding. 79% Of all quotes were coded as scientifically correct.

1. Introduction

Research on digital educational games shows promising results on potential learning benefits and their use in the classroom [1,2]. However, as shown in a large meta-analysis by Clark and colleagues [3], huge variances exist in terms of measured learning outcomes. This raises questions on what elements in the design of a game contribute to its efficacy and how such elements can be designed into the game. One of the most important aspects of the design process is the combination of learning elements within the gameplay, so-called intrinsic integration [4]. As shown in a more recent meta-analysis by Lamb et al. [1], implementing a specific pedagogical approach inside a game design is more effective than adding a pedagogical approach after the game has been developed. A third meta-analysis by Tsai and Tsai [2] found that gaming mechanisms and learning mechanisms are equally important in students' science learning whilst playing educational games. However, concrete examples of how to combine learning with gameplay remain scarce and therefore methodologies of combining learning approaches with gameplay remain largely unexplored [2,5].

### 1.1 Implementation of educational games

Apart from this relative dearth of studies on the combination of learning with gameplay, an educational game usually is part of a larger structure, such as a classroom-based lesson or an online synchronous meeting. Classical interventions such as Inquiry Based Learning, Direct Instruction or different teaching and learning activities are often combined into a larger whole. As an example, the effect of inquiry-based learning methods may depend on the amount of guidance provided in the lesson [6].

In order to increase the learning effects of educational games, it is important to consider how the game is embedded in the total set of learning activities [1,7]. Several studies suggest that a "debriefing session" in the form of a classroom discussion after the gameplay [7-9] will enhance learning when compared to gameplay alone. Studies also emphasize the teacher's critical role in connecting the game with other learning activities and other outside-game scenarios [9,10]. Other instructional activities, such as practical work or other assignments, can complement the game activities in a lesson and thus further foster learning and transfer.

One of the major reasons for embedding a game in a larger learning environment is that conceptual knowledge should also be applicable outside of the game context. In other words, students should be able to *transfer* their newly acquired knowledge from the game context to a wider one. For example, in the case of Newtonian mechanics, one learning goal is about understanding the relationship between forces and motion. Students should be able to apply this relationship on many different types of situations. If the game situation is, e.g, a rolling ball on a trajectory, students then should also be able to apply the new knowledge to other situations, such as riding a bike, moving a box or shooting a hockeypuck.

Students' ability to transfer knowledge from one situation to another depends on their understanding of the underlying principles and concepts. If students are able to construct an abstract representation of these underlying principles and apply this in other situations, transfer is fostered [11]. One possible way to achieve transfer is by abstraction and finding connections [12]. In terms of an educational game, this means that the specific game elements need to be taken to a more abstract level and connections with other parts of the educational program need to be made explicit. This points in the direction of embedding the educational game into a larger educational structure.

This case study focuses on learning and transfer of concepts in Newtonian mechanics when an educational game is embedded in a lesson. The game, Newton's race, is an example of a game in which intrinsic integration is applied [13]. In Newton's Race, players have to move a ball over a preset trajectory, with different levels for different trajectories.

The learning goal of the entire lesson is to understand and apply the relationship between forces and motion. In order to achieve transfer from the game situation to a wider context, three lesson activities are added after students play the game. The first activity, designed in order to increase learning and to abstract the acquired knowledge from the educational game, is a teacher-directed debriefing session directly after gameplay [see e.g., 9,10]. During this classroom discussion the teacher plays a central role by asking questions such as: 'What did you think you have learned from the game?' and 'can you explain what happens in this (game)scenario?' In the second activity, students are asked to construct an abstract representation of underlying principles [12]. Students get a generalization assignment asking them to write down what they have learned. In order to prompt students to make general statements about forces and motion, they are asked to use certain terms, such as acceleration, deceleration, forward force, resistance, bigger than, equal to, etc. The final activity is designed in order to apply these abstract general statements in different transfer situations [11]. Students are presented with different outside game scenarios, such as a rocket in space and riding a bike. Students then are asked to describe the type of motion in each scenario and to give an explanation for their chosen type of motion.

### 1.2 Research questions

A previous study on an educational game on Newtonian mechanics (Newton's Race) showed that the game resulted in a significant increase of domain knowledge presented *within* the context of the game. [14]. However, the posttest demonstrated that players were not able to transfer the acquired knowledge to other situations. As players only played the game for 15 minutes and as there were no additional instructional activities, this result may not come as a surprise. The question then becomes how to improve this transfer in cases where such an educational game is employed. Therefore, we raise the following research question:

*What is the impact on learning outcomes in terms of conceptual understanding and transfer of embedding an educational game in a lesson?*

This question is subdivided in the following sub-questions:

SQ1: To what extent do students understand the relationship between forces and motion in both game related and transfer situations after the lesson?

SQ2: What do students report on their understanding of the relationship between forces and motion in both game related and transfer situations?

Based on the findings of the aforementioned study [14] we hypothesize that by adding specific lesson activities, transfer can be fostered. Therefore, we expect a learning effect both in-game related and in transfer situations.

**1.3 Theoretical background, intrinsic integration**

Despite differences in exact terminology in studies on the combination of learning and gameplay, there is consensus on the main point: Learning and gameplay must be integrated during the design process in order to increase the learning effect of an educational game. This integration of subject matter with gameplay is defined by Kafai [4] as *intrinsic integration*. However, designing a game on the basis of this principle is not straightforward [15].

In an earlier study, we proposed a guiding frame for designing an intrinsically integrated game, shown in Figure 1. In this view, an intrinsically integrated game has four key elements that should be integrated; the learning goal, game goal, pedagogical approach and game mechanics. Integrating learning with gameplay means that the game goal is aligned as closely as possible with the learning goal (top part of Figure 1). In every game a player interacts with the environment through game mechanics in order to reach a certain game goal. Sicart [16] defines game mechanics as 'methods invoked by agents, designed for interaction with the game state'. This is represented in the left part of Figure 1. Turning to the right part, in order to reach any learning goal, a specifically chosen pedagogical approach is required. Ideally, in designing an intrinsically integrated game, the pedagogical approach needs to be aligned with the game mechanics.

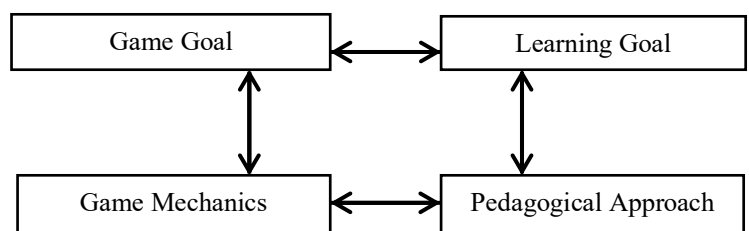

**Figure 1.** Guiding frame on alignment between the game goal, learning goal, pedagogical approach and game mechanics for designing an intrinsically integrated educational game, used for designing Newton's Race [13].

**1.4 Pedagogical approach and learning goal**

For this study the intrinsically integrated game Newton's Race was used. Using the terminology from Figure 1 the design of Newton's Race, in order to establish the intrinsic integration, can be described. A more detailed version of the design process can be found here [13].

The subject context of Newton's Race is Newtonian mechanics. The chosen *learning goal* for the game is that students can reason about the effects of forces on different types of motion. This implies that students can understand and apply the relationship between forces and motion. Students reach the learning goal when they understand which force (no force, a force equal to friction or a force bigger than friction) corresponds to a certain type of motion: (acceleration, deceleration or constant velocity).

For intrinsic integration to apply, a specific, domain related *pedagogical approach* is needed that is integrated with game mechanics [1,15 and Figure 1]. The pedagogical approach chosen for the game is based on the problem posing approach. This approach has shown promising results on several topics in physics education [17-19]. Before encountering Newtonian mechanics in formal education, students all have gathered daily experiences with moving objects, resulting in pre-existing ideas about this subject. In order to accept a new scientific explanation on these subjects, students need to be confronted with the value of their pre-existing ideas [20-22]. Ideally, when students are confronted with the consequences of their pre-existing ideas, they experience something that counters their expectation and they are likely to experience a need to alter their own explanation to a new, scientific one.

In order to align the *game mechanics* with this pedagogical approach, two main concepts must be present in the game. First, players should be able to explore their pre-existing ideas. To accommodate this, a setting phase is required were players can set different forces on a moving object. Thus, players are confronted with the consequences of their pre-existing ideas. Therefore, the game mechanics must take into account the experience of the type of motion as a consequence of the setting phase.

Turning now to the game goal and game mechanics, the *game goal* is guiding a ball to the finish line of a preset trajectory. The integration of the game and learning goals is reflected in Newton's Race consisting of five levels, with each level designed in such a way that players can only reach the finish line, if their ball moves in a realistic fashion, thus using the scientifically correct setting in the setting phase.

The details on the gameplay can be found in an earlier publication [14]. As an illustration some screenshots of the game are included in Figure 2.

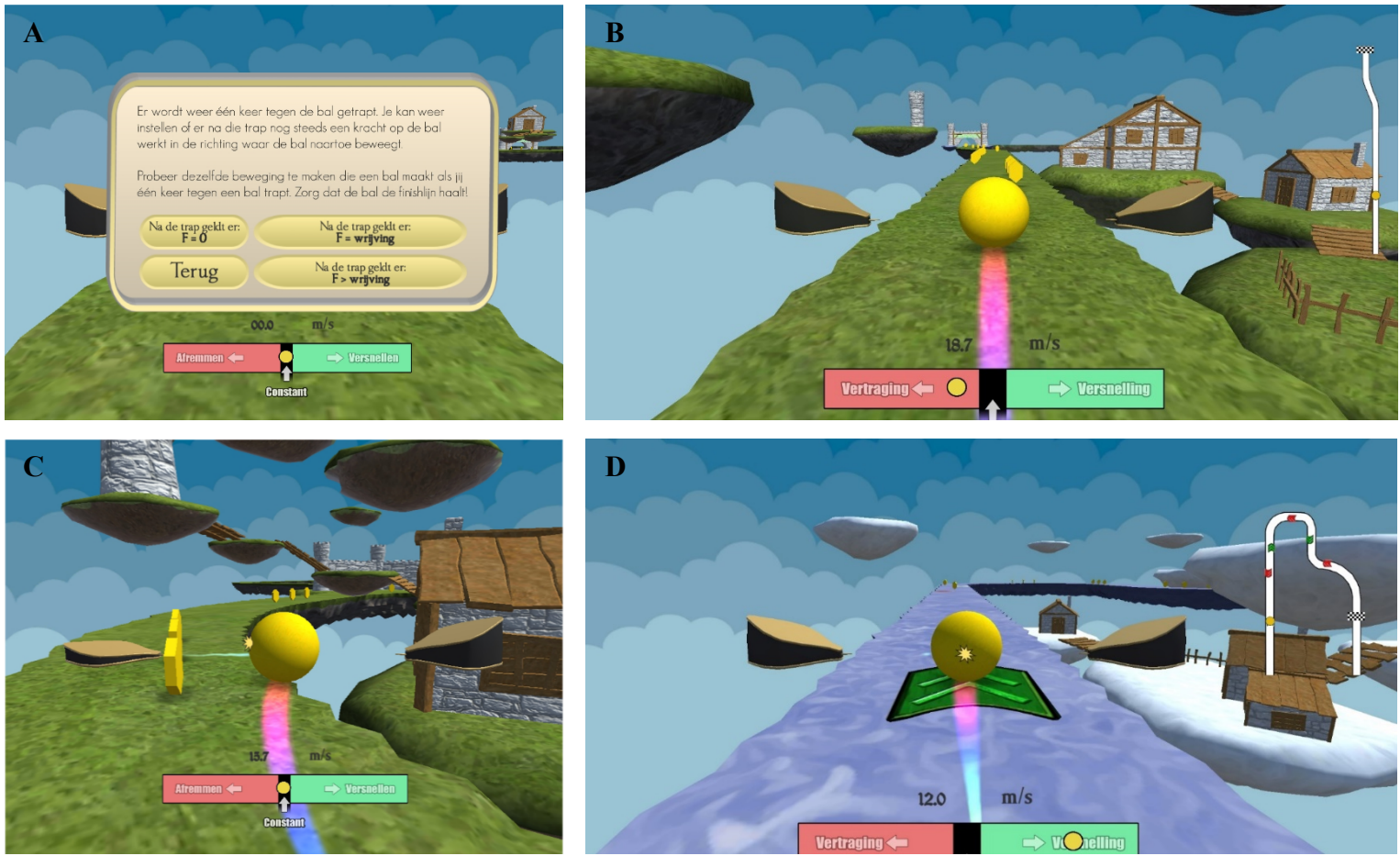

**Figure 2. A.** The setting phase at the start of each level. Translated: A ball is kicked once. You can set if there is a force working on the ball in the direction of motion. Try to make the same movement a real ball makes when you kick it once. Make sure the ball gets to the finish line! Options: 'After the kick: F = 0', 'After the kick: F = friction', 'After the kick: F > friction' or 'Go back'. **B.** Gameplay of level 2 of Newon's Race. **C.** Changing the direction of motion of the ball in order to collect coins and to take turns on the trajectory. **D.** On an acceleration platform (green) a force is applied in the direction of motion. On a deceleration platform (red) a force is applied in the opposite side of the direction of motion.

2. **Materials and methods**

**2.1 Research design**

For this study a mixed methods approach was used. Students participated in a lesson about Newtonian mechanics. The lesson plan used for this study can be found in Table I. A pre- and posttest design was used to examine what students learned about the relationship between forces and motion (SQ1). Audio data and worksheets were collected in order to qualitatively examine the students' understanding of this relationship (SQ2). The lesson plan was piloted in a 9[th] grade physics class in a medium-sized urban secondary school in the Netherlands, after which minor procedural adjustments were made for the final study.

**2.2 Participants**

Participants were selected on a voluntary basis from eight different classes in the same school and consisted of Dutch 9$^{th}$ grade students between the age of 14 and 15. They participated in the study in four groups of six, seven, and eight students, respectively. From a total of 27 participants, 21 (78%) were boys and 6 (22%) were girls. Since the division into groups was done arbitrarily on the basis of the availability of the students, results from all three groups will be lumped and no distinction between the groups will be made.

**2.3 Instruments**

The pretest consisted of five questions about situations closely related to the game situation. These five questions were matched with the five same questions in the posttest to determine a possible learning effect. The posttest contained three additional questions regarding situations not closely related to the game, to determine a possible transfer effect (SQ1) as is shown in Figure 3. The questions used in the pre- and posttest are based on questions of the Force Concept Inventory [23] and Appendices A and B gives the exact pre- and posttest questions.

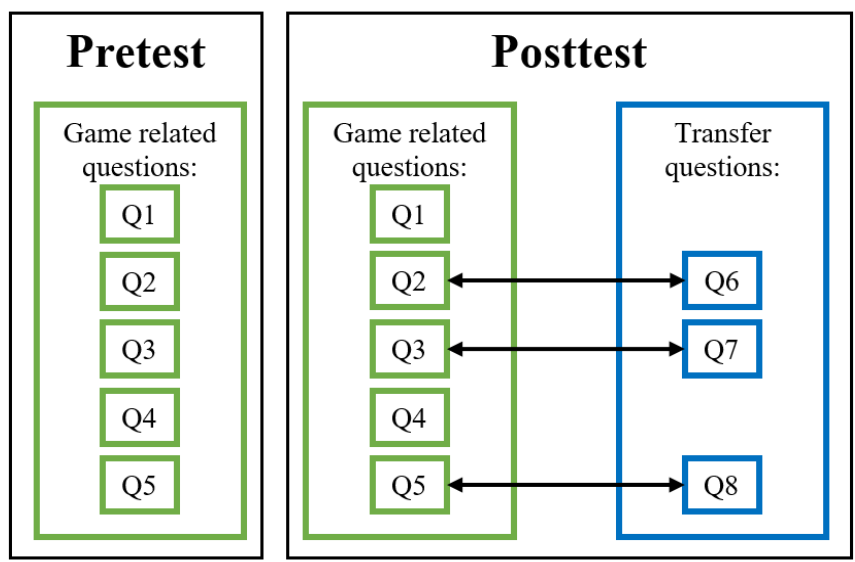

**Figure 3.** Visual representation of the pre- and posttest questions, showing the conceptual connections between the questions. E.g., Q2 and Q3 deal with the same concept, Q3 addresses the concept in a game related context and Q6 addresses the same concept in a different context. The complete pre- and posttest can be found in Appendices A and B.

**Table I:** The lesson plan used for the study.

| Duration (min) | Activity | Short description | Data collection |
|---|---|---|---|
| 3 | 1. Introduction | | |
| 7 | 2. Pretest | | Pretest |
| 5 | 3. Discussion on statement 1 | Activation of pre-existing ideas by a discussion on an everyday life scenario: riding a bike in bad weather. | Audio data |
| 15 | 4. Playing Newton's Race | Students play the game individually. | |
| 15 | 5. Classroom discussion | A debriefing session consisting of three parts: 1. Gameplay and visual representations 2. What did you learn from the game? 3. Discussion based on screenshots of the game | Audio data |
| 15 | 6. Worksheet | Students fill in a worksheet consisting of 2 assignments: 1. Generalization assignment 2. Appling the generalization to transfer situations | Worksheet & Audio data |
| 5 | 7. Discussion on statement 2 | As a closing activity: another discussion on an everyday life scenario: moving a heavy box. | Audio data |
| 10 | 8. Posttest | | Posttest |

Table I gives an overview of the lesson plan. After the introduction and pretest, students pre-existing ideas were activated by a discussion on an everyday life scenario: riding a bike in bad weather (3). Then, students were invited to play the educational game Newton's Race (4). Directly after playing the game, a classroom discussion consisting of three parts was held (5). The first part was about gameplay and visual representations. This part was primarily used for priming the discussion and to get a feeling for the way the game "handled". The results of this part of the study will not be discussed here, since they are not related to the research questions. In the second part the following general question was raised: What did you learn from the game? In the third part screenshots of Newton's Race were used to spark the discussion. Various in-game scenarios were shown on a handout, students were asked to discuss what happened to the ball on a particular moment, how they figured what type of motion the ball was in, and what kind of setting would be the correct one for that motion.

After this debriefing discussion on the game, students individually filled out a worksheet (activity 6) with generalization assignments, which was also discussed thereafter. Due to the additional discussion element of activity 6, some overlap in statements will occur between the individually filled out worksheets and the audio data. The worksheet developed for this lesson consists of two assignments designed to foster transfer. In the first assignment students were asked to generalize the acquired knowledge from the game and classroom discussion. In the second assignment students had to use that generalization in order to explain the relationship between forces and motion in five different transfer situations. At the end of the lesson, another short discussion on a statement followed, this time about pushing a heavy box (7).

The pretest, posttest, statements and the worksheet, translated from Dutch, can be found in respectively Appendices A, B, C and D.

### 2.4 Data collection and analysis

The duration of the lesson was 75 minutes, including the pre- and posttest. General data, such as age and school grade, as well as pre- and posttest data were collected digitally using Google Forms. In order to answer SQ1, a paired sample *t*-test was performed on the pre- and posttest scores, in order to measure the direct learning effect. A paired sample *t*-test was also performed on the transfer questions in order to determine a possible transfer effect from game related situations to other situations.

In order to answer SQ2: The audio data of the discussions on the statements and of the second and third part of the classroom discussion were transcribed verbatim. The worksheets filled out by the students were collected anonymously. Open coding was used on both the transcripts and the worksheets in order to group statements related to forces and motion. Several categories were found in order to describe what the statement is about: A (acceleration), D (deceleration), C (constant velocity), G (General or other statement about motion) and F (force). Multiple codes could be assigned to a statement, e.g., in the case a statement on the relationship between forces and motion. Statements were also coded on the scientific correctness of that statement using a + or – sign. To gain more insight in the coding process, Table II presents the coding scheme and Table III gives some coded examples. Two interrater reliability analysis were performed on 30% of the collected statements. For the different categories a Cohen's Kappa of .994 ($p$ = .006) was found. For the correctness of statements a Cohen's Kappa of .989 ($p$ = .011) was found.

**Table II:** The coding scheme used to code the transcribed statements. The coding scheme is translated from Dutch. In addition to the code categories, a + or – is assigned depending on the correctness of the statement.

| Code | Category | Examples of statements |
|---|---|---|
| F | Force | I think F = 0 / It receives a force from the left / Resistance / If the forward force is greater than the opposing force |
| A | Acceleration | You see an acceleration / The ball is going faster and faster |
| D | Deceleration | You see a deceleration / The ball is slowing down / He stops |
| C | Constant velocity | You move with a constant velocity / constant speed |
| G | General or other statement about motion | You go forward / The ball is rolling / One goes twice as fast / The ball stands still |

**Table III:** Some examples from coded statements, translated from Dutch.

| Lesson activity | Statement | Codes |
|---|---|---|
| 3. Discussion on statement 1 | If all forces are the same, you just stand still | F-, G- |
| 5. Classroom discussion | At the bottom one, there is still a force so it will probably accelerate the whole trajectory | F+, A+ |
| 6. Worksheet assignment 2 | Because ice offers almost no resistance, it will remain at a constant velocity for a long time | F+, C- |

3. **Results**

**3.1 SQ1: Conceptual understanding**

The first sub-question was: *To what extent do students understand the relationship between forces and motion in both game related and transfer situations?*

Table IV gives the results of the pre- and posttests for game-related and transfer situations.

**Table IV:** Results of the pre- and posttests (with a minimal value of 0 and a maximal value of 5 for the game related questions and a maximal value of 3 for the transfer questions).

|           | Game related |          | Transfer |          |
|-----------|--------------|----------|----------|----------|
|           | Pretest      | Posttest | Posttest | Posttest |
| Questions | 1,2,3,4,5    | 1,2,3,4,5 | 2,3,5   | 6,7,8    |
| Mean      | 2.89         | 3.93     | 2.63     | 2.56     |
| SD        | 1.15         | 1.14     | .63      | .64      |

A paired samples *t*-test was performed to examine the mean differences between the questions on game related situations in the pretests (M=2.89, SD=1.15) and the posttests (M=3.93, SD=1.14). A significant difference was found; $t(26) = -3.463$, $p = .002$, Cohen's $d = .908$. Another paired samples *t*-test was performed to examine the mean differences between posttest questions on game related situations (M=2.63, SD= .63) and transfer situations (M=2.56, SD= .64), No significant difference was found; $t(26) = 1.000$, $p = .327$.

### 3.2 SQ2: Conceptual understanding

The second sub-question was: *What do students report on their understanding of the relationship between forces and motion in both game related and transfer situations?*

We start this section by presenting in Figure 4 the number of total codes per category. A distinction is made between the type of collected data: audio data or the individually filled out worksheets.

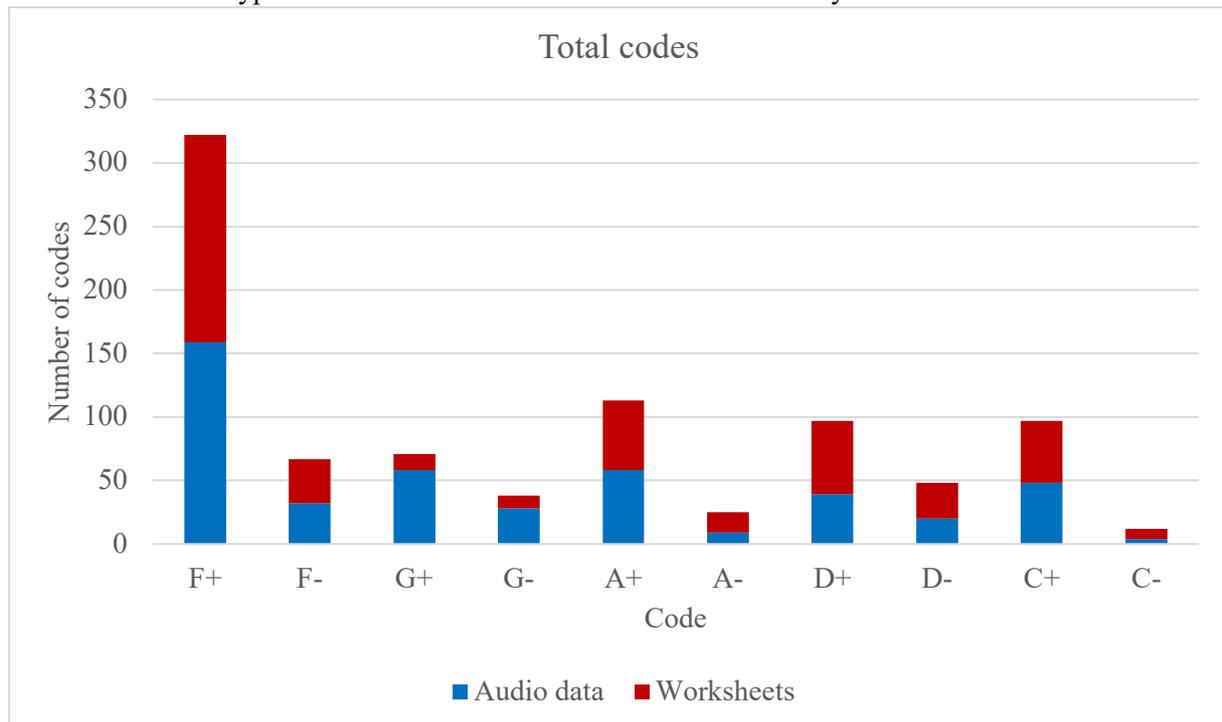

**Figure 4:** In this graph the number of codes is given per category. The categories are statements regarding: force (F), acceleration (A), deceleration (D), constant velocity (C) and other type of movements (G). Statements in each category were also coded scientifically correct (+) or incorrect (-).

Looking at Figure 4, several aspects stand out.

1) Most statements were coded within the F+ category, 36% of all codes.

2) More scientifically correct codes (+) were given than incorrect codes (-), 79% of all codes were scientifically correct.

3) Incorrect statements on acceleration, deceleration and constant velocity (A-, D-, C-) were primarily given on the worksheets (64%, vs 36% on the audio data).

4) Statements on general or other type of movements (G) were primarily given (79%) on the audio data.

### 3.2.1 Game related situations

In the second part of the discussion (5.2 in Table I) students are asked to describe what they think they have learned from the game. In each group something came up on forces and motion in an attempt to generalize what they experienced in the game. The following short excerpt from a transcript of a group discussion illustrates how a student responds to the question 'What did you learn from the game':

R(esearcher): *Can anybody tell in their own words if they learned anting from the game and if so, what did you learn?*

S(tudent): *What a force does to a ball.*

R: *OK and what does a force do to a ball?*

S: *Well, if F = 0, then he will slow down (ehr), the F is bigger, the force is bigger than friction, than it will keep getting faster.*

This statement shows that this student is referring to the game, by referring to 'F=0' (which is a game setting) and 'ball'. This student in particular shows an understanding in game related situations of the relationship between acceleration and forces and deceleration and forces. In the last sentence two correct relationships are given (F+, D+ and F+, A+), however no statement is made on the relationship between force and moving with a constant velocity.

In the last part of the discussion (5.3 in Table I) the students were given a three-page handout with several screenshots from the game. On the first page the students were asked to tell the differences between two screenshots of the first level, a trajectory over ice. The screenshots differ as they represent different settings resulting in (1) a difference in type of motion (a constant velocity or acceleration) and (2) difference in force working on the ball. All groups were able to point out these two differences and were able also to explain *how* they could tell that there was a difference in motion (for instance by reading the accelerometer). The following excerpts from transcripts illustrates typical students responses on the difference between these two screenshots:

S: *At the bottom one there is still a force, so it will accelerate the rest of the track* (F+, A+)

R: *How can you tell that there is a force working?*

S: *Because you see a point at the middle of the ball* [See Figure 2D]

S: *With that one the dot is in the accelerated section and with the other it is in the constant section and one gets a force and the other doesn't.* (F+, A+, C+)

On the second page of the handout students had to tell de differences between two screenshots on ice and on grass, the ball moving with a constant velocity in both situations. The two important differences to be found here were: 1) the difference of the track (no friction vs. friction) and 2) the difference in force. Again all groups were able to spot these differences. Students statements also show that although the type of motion (constant velocity) is the same, the set force must be different in the two levels. The following two excerpts from a transcript illustrates students responses on the difference between these two screenshots:

S: *The top one is blue and the bottom one is green*

R: *And what does it mean that the top one is blue and the bottom one green?*

S: *I assume that the blue represents ice, so there is less friction than green, the grass.* (F+)

R: *Euhm I think that with the grass one that there an extra force must be applied on, because otherwise the grass will slow it down.* (F+, D+)

R: *OK and what does that mean for the top picture?*

S: *That there it is not the case*

On the third page three screenshots were shown, all with the ball moving on grass. The students were asked what kind of motion they saw and what force was set to create that motion. All groups could quickly state the correct type of motion that they saw in the screenshots and they were also able to reproduce the setting that was used to create that motion.

### 3.2.2 Transfer situations

In the first assignment of the worksheet (6.1 in Table I) students had to generalise their understanding on the relationship between forces and motion. Ideally, they were supposed to come up with three physically correct relationships:

1. When the applied force is bigger than friction, the object is accelerating.
2. When the applied force is smaller than friction, the object is decelerating.
3. When the applied force is equal to friction, the object is moving with a constant velocity.

As is shown in Table V, 48% of the students produced all the above relationships and 11% produced none of the relationships. From the 41% that partially produced the relationships, 45% produced relationship 1, 45% produces relationship 2 and 55% produced relationship 3.

**Table V:** Results of the generalization assignment.

|  | N | % |
|---|---|---|
| All relationships | 13 | 48 |
| Part of the relationships | 11 | 41 |
| No relationships | 3 | 11 |

In the second assignment of the worksheet students had to apply their understanding on the three mentioned relationships in five different transfer situations. For each statement students had to give the correct motion and explain their answer by using the relationships. Each correct answer was coded with half a point and half a point for the correct explanation, so a maximum of 5 points could be gathered here. The actual results can be found in Table VI. NB: These results are independent of the transfer questions in the posttest.

**Table VI:** Results of the second worksheet assignment (with a minimum value of 0 and a maximum value of 5). The students are divided into three groups, dependent of their answers on the generalization assignment.

|  | N | Mean (out of 5) | SD |
|---|---|---|---|
| All relationships | 13 | 3.54 | .901 |
| Part of the relationships | 11 | 2.81 | 1.15 |
| No relationships | 3 | 2.83 | .289 |
| Total | 27 | 3.17 | 1.01 |

There was no statistical significant difference between the answers of different groups as determined by a one-way ANOVA ($F(2,24) = 1.806$, $p = .186$).

## 4. Conclusions and discussion

### 4.1 Conclusions

Revisiting our sub questions:

SQ1: *To what extent do students understand the relationship between forces and motion in both game related and transfer situations?*

Results show a large significant learning effect ($d = .908$) on the game related questions, confirming results of our hypothesis and earlier study [14]. Also confirming our hypothesis, transfer was fostered when the game is embedded in a lesson. This is based on the results showing no significant difference between the questions in the posttest on game related situations (Q2,3,5) and transfer situations (Q6,7,8 in Figure 3). In other words: after the complete lesson, students were able to apply their acquired knowledge with equal success in both game related and transfer situations. These findings indicate that after this lesson students' conceptual understanding on the relationship between forces and motion improved significantly in both game related and transfer situations.

SQ2: *What do students report on their understanding of the relationship between forces and motion in both game related and transfer situations?*

As is shown in Figure 4, students most often used the concept of force in their utterances (both written and oral) compared to the other categories that are all about motion. Since the learning goal is about the relationship between forces and motion, it is to be expected that for each utterance on a type of motion (acceleration, deceleration or constant velocity) students also report something about the forces involved. The concept of force is thus expected to be used more often than the other categories.

Students statements were mostly scientifically correct (79%) indicating students overall ability to reason about forces and motion. The incorrect statements that were specifically made about acceleration, deceleration and constant velocity (A-, D-, C-) were primarily (64%) given on the worksheets, which were filled out individually. This could be interpreted in two ways. First, if a student is insecure about their answer, they could be inclined not to give that answer in a classroom discussion and remain silent. On the worksheet however, every student is expected to give an answer. Another explanation could be that on the worksheet students were explicitly prompted to use acceleration, deceleration and constant velocity in their answers. This is also in line with the finding that the general statements about motion (G) were mostly given during the discussions and not on the worksheets (79%).

During the debriefing session directly after students played the game, all groups proved to be capable to reason on the relationship between forces and motion in game related situations. During the discussion on screenshots from the game, all groups were able to interpreted the situations correctly. When asked the question 'what did you learn', students report scientifically correct, but sometimes incomplete, relationships between forces and motion.

In the generalization assignment of the worksheet, about half (48%) of the students were able to give a complete, and scientifically correct, set of statements on the relationship between forces and motion. Most of the other students (41%) gave correct, however incomplete set of statements, and only 11% of the students gave no correct statements at all. When comparing these groups in their results on the second worksheet assignment (with transfer situations), no significant differences in mean scores were found. Therefore, a better score on the generalization assignment did not translate into a significantly higher score on the assignment with transfer situations.

We now revisit the main research question: *What is the impact on learning outcomes in terms of conceptual understanding and transfer of embedding an educational game in a lesson?*
The findings of this case study indicate that students profit from a debriefing session after playing an educational game. During this debriefing session students demonstrate their understanding of the relationship between forces and motion in game related situations. Although no correlation is found between the scores of the two worksheet assignments on generalisation and transfer, we did find a learning and transfer effect in posttest scores. In comparison with the previous study [14], the effect size was larger ($d$ =.201 vs. $d$ = .908) and in the present study, transfer was fostered as well. We can thus conclude that the additional learning activities surrounding Newton's Race used in this study increased the learning effect and fostered transfer.

### 4.2 Situating the study

In an attempt to further close the gap on how to effectively implement an educational game in a lesson, this case study on Newton's Race shows positive results by explicitly describing and analysing lesson activities surrounding the intrinsically integrated game Newton's Race. Our findings are in line with our hypothesis based on a previous study [14] and with other studies. The positive effect of the lesson as a whole is in line with studies on three lesson activities after playing Newton's Race. The three lesson activities being: a teacher directed debriefing session [7,9,10], a generalization assignment [12] and an assignment on different transfer situations [11].

### 4.3 Limitations and implications

In interpreting the results of the current study some limitations must be taken into account. Firstly, participants were recruited on a voluntary basis, as a result a selection bias may have occurred, more boys than girls participated in this study. Participants may have had a positive attitude towards games in general before participating in the study. Secondly, the sample size was quite small (27 students). Thirdly, when interpreting the results of the worksheets one must keep in mind that the worksheet was discussed in a group discussion only *after* the students filled them out individually. Therefore, some overlap in coding inevitably occurred between the transcripts of the audio data of that discussion and the worksheets. Also, students could gain new insights during that discussion that was not reflected in their worksheets. Finally, no retention test was conducted. This study is meant to give an insight in how transfer can be fostered with regards to an intrinsically integrated game. Research on a possible retention effect is the next step in the design process.

The findings of this study show the importance of embedding educational games in lessons in order to foster transfer. When designing a lesson were an educational game is used, the designer must also take additional activities into account, such as a debriefing session, a generalization assignment and an assignment on different transfer situations.

### 4.4 Further research

In this study a lesson is evaluated where no feedback from the teacher was given regarding answers being right or wrong or any type of explanation on the relationship between forces and motion, even if the students explicitly asked for it. This was necessary for this study in order to gain a true insight on students' reasoning without their teacher's feedback on how they were doing. In reality, of course, the importance of the role of (teachers) feedback on learning has been proven time and time again [e.g., 24] When implementing this lesson in the actual classroom, there will of course be teacher's feedback at the end of a discussion, affirming the understanding of students' relationship between forces and motion, especially in transfer situations.

The findings of this case study show some promising directions on how to foster transfer using an intrinsically integrated game on Newtonian mechanics in a lesson situation. The next step is to expand these findings by evaluating a lesson around Newton's Race that fosters transfer in a real classroom situation. For this to happen, the developed lesson needs to be adjusted accordingly. After all, teaching eight students (the maximum number of students in one of the four groups in this case study) is not the same as teaching 30 students in a full class. Also, research on a possible retention effect is needed to further evaluate the effects of a lesson where an intrinsically integrated game, such as Newton's Race, is implemented.

**Acknowledgements**

The present article was made possible by funding from the Dutch Ministry of Education, Culture and Science, OCW/PromoDoc/1065001. We would also like to thank Wouter Korff and Chrisjan Kottelenberg for their contribution to the programming of Newton's Race.

Conflict of Interest: The authors declare that thy have no conflict of interest.

Ethics approval: This study was performed in line with the ethical standards of the Freudenthal Institute. Informed consent was obtained from all individual participants included in the study.

**Appendix A: Pretest (translated from original language)**

1. By kicking a stationary ball, the ball starts moving. Then there will be a force working on the ball in the direction of motion.
True / False / I do not know

2. You kick a ball through long grass. When you stop kicking, the ball will stop moving immediately.
True / False / I do not know

3. You kick a ball once, so that it rolls over grass. Does a force work on the ball in the direction of motion after the kick?
After the kick:
A. There is no force working.
B. There is a force working. The force is equal to opposing forces (from the grass).
C. There is a force working. The force is bigger than opposing forces (from the grass).
D. I do not know.

4. You kick a ball once, so that it will roll across grass. What kind of motion will the ball make?
A. First faster and then slower.
B. Ever slower.
C. First slower and then faster.
D. Ever faster.
E. A constant speed.
F. I do not know.

5. You kick a ball across grass, so that the ball moves with a constant speed. The forwards kick-force on the ball is:
A. Bigger than opposing forces (from for instance the grass).
B. Smaller than opposing forces (from for instance the grass).
C. Equal to opposing forces (from for instance the grass).
D. I do not know.

**Appendix B: Posttest (translated from original language, number of questions differ slightly from figure 3)**

1. By kicking a stationary ball, the ball starts moving. Then there will be a force working on the ball in the direction of motion.
True / False / I do not know

2. You kick a ball through long grass. When you stop kicking, the ball will stop moving immediately.
True / False / I do not know

3. You slide a couch across the floor. When you stop pushing, the couch will stop moving immediately.
True / False / I do not know

4. You kick a ball once, so that it rolls over grass. Does a force work on the ball in the direction of motion after the kick?
After the kick:
A. There is no force working.
B. There is a force working. The force is equal to opposing forces (from the grass).
C. There is a force working. The force is bigger than opposing forces (from the grass).
D. I do not know.

5. You are riding a bike and at a certain moment you stop pedaling. Does a force work the direction of motion after you stop pedaling?
After the kick:
A. There is no force working.
B. There is a force working. The force is equal to opposing forces.
C. There is a force working. The force is bigger than opposing forces.
D. I do not know.

6. You kick a ball once, so that it will roll across grass. What kind of motion will the ball make?
A. First faster and then slower.
B. Ever slower.
C. First slower and then faster.
D. Ever faster.
E. A constant speed.
F. I do not know.

7. You kick a ball across grass, so that the ball moves with a constant speed. The forwards kick-force on the ball is:

A. Bigger than opposing forces (from for instance the grass).

B. Smaller than opposing forces (from for instance the grass).

C. Equal to opposing forces (from for instance the grass).

D. I do not know.

8. You roll a hockey ball across grass, so that the ball moves with a constant speed. The forwards roll-force on the ball is:

A. Bigger than opposing forces (from for instance the grass).

B. Smaller than opposing forces (from for instance the grass).

C. Equal to opposing forces (from for instance the grass).

D. I do not know.

**Appendix C: Statements (translated from original language)**

Statement 1:

You ride your bike to school and there is a lot of headwind and rain. Because of that you always ride with 14 km/h to school. The following applies for your pedaling force:
1. The pedaling force is 0 N.
2. The pedaling force is smaller than the opposing forces (for example by the wind).
3. The pedaling force is equal to the opposing forces (for example by the wind).
4. The pedaling force is bigger than the opposing forces (for example by the wind).
5. I do not know.

Statement 2:

You are moving and you slide a heavy box across the floor. When you stop pushing, the box will slide a bit further. The following applies:
1. There is no fore working in the direction of motion.
2. There is a fore working in the direction of motion. This force is smaller than the opposing forces (for example by the floor).
3. There is a fore working in the direction of motion. This force is equal to the opposing forces (for example by the floor).
4. There is a fore working in the direction of motion. This force is bigger than the opposing forces (for example by the floor).
5. I do not know.

**Appendix D: Worksheet (translated form original language)**

1. Write down for yourself what you have learned. Use the following terms:
forward force; acceleration; bigger than; deceleration; smaller than; opposing force; equal to; constant velocity.
You may use the terms more than once.

2. Indicate for the situations below what kind of motion applies. Give an explanation for the chosen motion, using your answer form task 1.

a. A rocket turns off its thrusters in space.
b. A hockey puck is being hit by a hockey stick.
c. After a hockey ball has been hit, it continues to move across grass.
d. After an ice hockey puck is hit, it moves on over ice.
e. You cycle down a hill, without pedaling you roll down.